\documentstyle[aps,prl,twocolumn,floats,psfig]{revtex}

\begin{document}
\psfigurepath{.:plot:figure}

\narrowtext

\noindent {\bf Comment on ``Low-Energy Magnetic Excitations 
of the Mn$_{12}$-Acetate 
Spin Cluster Observed by Neutron Scattering''}

\vskip 2ex

The Mn$_{12}$-acetate (Mn$_{12}$) molecule is used  
to study the delicate interplay
between thermal-activation and macroscopic-quantum-tunneling relaxation 
processes\cite{sessoli,jrf}. The intramolecular exchange 
interactions
render it an effective $S=10$ nanomagnet at low temperature\cite{s10}.
The energy barrier that separates the $m>0$ 
and $m<0$ states has been experimentally determined to derive from the 
Hamiltonian
\begin{equation}
H_0=-A S^2_z-B S^4_z,
\end{equation}
where $A= 0.38(1)$ cm$^{-1}$ and $B= 8.2(2)\times10^{-4}$ cm$^{-1}$ 
\cite{barra,mukhin,mhc}. 
In order for there to be tunneling, the spin Hamiltonian must contain terms 
that do not 
commute with $S_z$.  In the absence of a magnetic field, the lowest order such 
term allowed by tetragonal symmetry is
\begin{equation}
H_1=\frac{1}{2} C (S^4_+ + S^4_-).
\end{equation}
This term is 
invoked to explain an enhanced magnetization relaxation rate when 
an external longitudinal magnetic field causes energy levels to 
cross\cite{fort}.
This is due to the fact that quantum tunneling between, say, the $m=4$
and $-4$ states provides a shortcut in the activation process so that the
activation energy is at the $|m|=4$ level instead of at the top of 
the energy barrier\cite{fort}. Therefore,
the effective activation barrier, thus the magnetization relaxation rate,
depends crucially on the value of $C$.

$C$ has been estimated from an EPR experiment to be
$4(1)\times 10^{-5}$ cm$^{-1}$ for Mn$_{12}$\cite{barra}. 
An inelastic neutron scattering (INS) experiment
reported in\cite{mhc} revised the value of $C$ downward to 
$3.0(5)\times 10^{-5}$ cm$^{-1}$.
However, this INS experiment did not have sufficient data
to make an accurate determination of $C$. In this Comment,
we show from our own INS data that $C$ is much smaller than both reported 
values.

A deuterated Mn$_{12}$ polycrystalline sample inside a
$30\times 50 \times 3$~mm$^3$ aluminum container was measured
with fixed incident energy $E_i = 1.45$~meV on IN5 at ILL. 
The sample temperature 
was held at 20.7~K to populate
the upper levels. Spin excitations between levels of different $m$
show up as peaks
in Fig.~1. 
\begin{figure}[ht]
\centerline{
\psfig{file=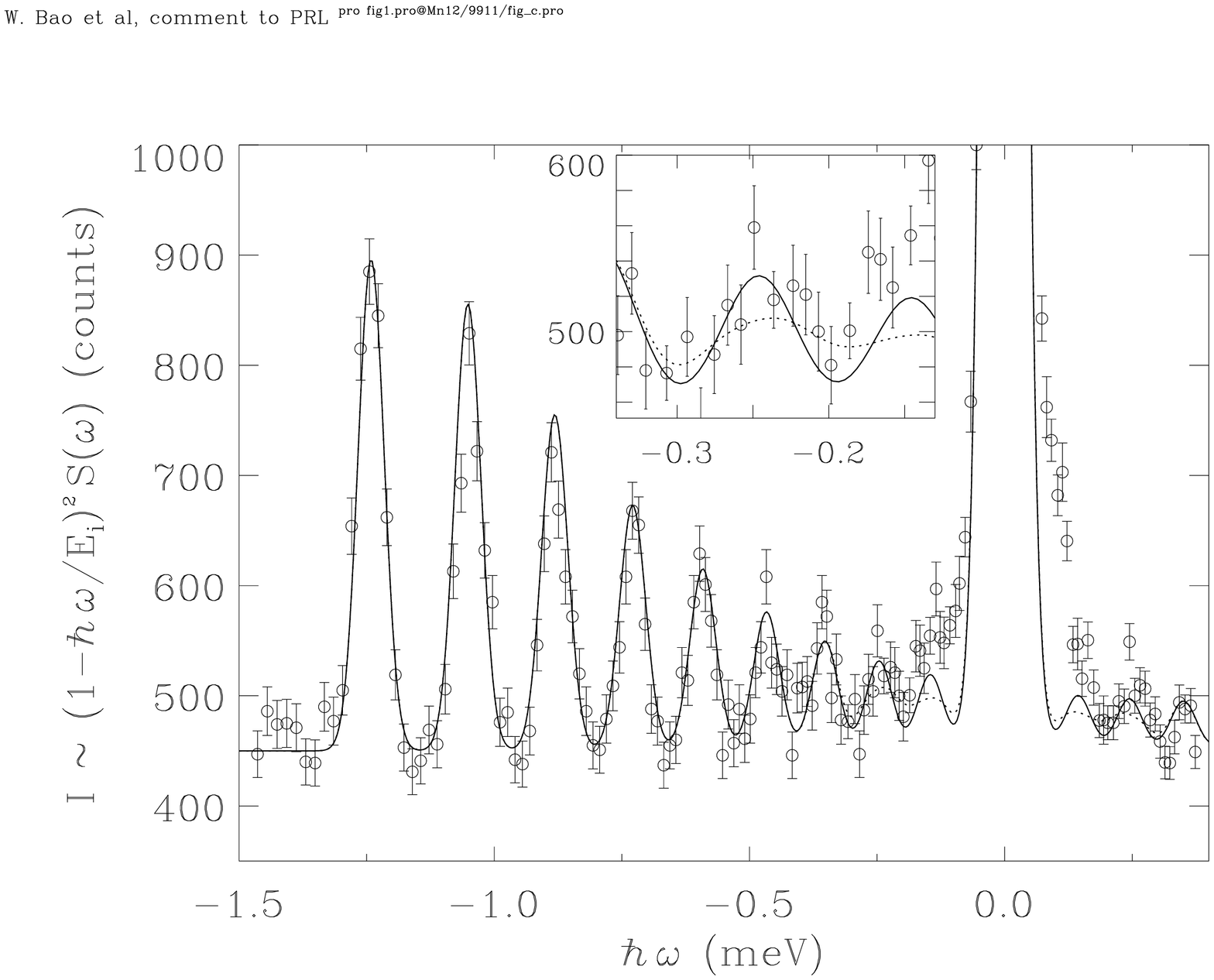,width=.5\textwidth,angle=90,clip=}}
\caption{Inelastic neutron scattering spectrum.  
The solid (dashed) curves are calculated spectra 
for C = 0 ($3.0\times 10^{-5}$ cm$^{-1}$), plus a constant background.
Extra intensity near $\hbar\omega=0$ is spurious\protect\cite{mhc}.  
The inset shows the low-energy 
part of the spectrum.}
\end{figure}
The solid and dashed curves are the calculated intensities,
convoluted with the experimental resolution, 
for $C=0$ and $C=3.0\times 10^{-5}$ cm$^{-1}$, respectively. 
The difference is mainly in the small energy-transfer region where
excitations between small $m$ levels occur.
Clearly, the $C=0$ curve fits our data better. The broadening and
shifting of the two lowest-energy peaks predicted by
a $C$ as large as $3.0(5)\times 10^{-5}$ cm$^{-1}$ is not
supported by our higher-resolution data.  From a careful analysis of our data
we have determined an upper bound for $\left|C\right|$ 
as $2.0\times 10^{-5}$ cm$^{-1}$.
The $H_1$ term, if it exists at all, 
therefore must be much smaller than has been reported previously.
Our data also yield $A=0.3855(8)$ cm$^{-1}$ 
and $B=7.80(6)\times10^{-4}$ cm$^{-1}$, which are consistent with the 
values reported in\cite{mhc}.

The values of $C$ reported in\cite{barra,mhc} have been used in 
some of the theoretical treatments of the magnetic relaxation 
of Mn$_{12}$\cite{fort}. 
Our experimental results may imply that other mechanisms, 
such as hyperfine fields\cite{jrf} or transverse external fields, 
may be more important
than $H_1$ in driving the tunneling.

Work at LANL is supported by US DOE.

\vskip 2ex

\noindent W.\ Bao$^{1}$, R.\ A.\ Robinson$^{1}$, J.\ R.\ Friedman$^{2}$, 
H.\ Casalta$^{3}$, E.\ Rumberger$^4$ and D.\ N.\ Hendrickson$^{4}$

$^{1}${\it Los Alamos National Laboratory, Los Alamos, NM 87544, USA}


$^{2}${\it The State University of New 
York, Stony Brook, NY 11794, USA}

$^{3}${\it Institut Laue Langevin, BP 156, F-38042 Grenoble, France}

$^{4}${\it University of California at San Diego, La Jolla, CA 92093, USA}

\end{document}